\documentclass{aa}  

\usepackage{graphicx}
\usepackage{txfonts}
\usepackage{natbib} 
\bibpunct{(}{)}{;}{a}{}{,} 

\usepackage{verbatim}
\usepackage{url}
\usepackage{epstopdf}
\usepackage{slashbox}
\usepackage{color}

\begin{document}

   \title{Calibrating convective-core overshooting with eclipsing binary systems} 

   \subtitle{The case of low-mass main-sequence stars.}

   \author{G. Valle \inst{1,2,3}, M. Dell'Omodarme \inst{3}, P.G. Prada Moroni
     \inst{2,3}, S. Degl'Innocenti \inst{2,3} 
          }

   \authorrunning{Valle, G. et al.}

   \institute{
INAF - Osservatorio Astronomico di Collurania, Via Maggini, I-64100, Teramo, Italy 
\and
 INFN,
 Sezione di Pisa, Largo Pontecorvo 3, I-56127, Pisa, Italy
\and
Dipartimento di Fisica "Enrico Fermi'',
Universit\`a di Pisa, Largo Pontecorvo 3, I-56127, Pisa, Italy
 }

   \offprints{G. Valle, valle@df.unipi.it}

   \date{Received 17/09/2015; accepted 21/12/2015}

  \abstract
{Double-lined eclipsing binaries have often been adopted in literature to calibrate the extension of the convective-core overshooting beyond the border defined by the Schwarzschild criterion.}
   {  
In a robust statistical way, we quantify  the magnitude of the uncertainty that affects the calibration of the overshooting efficiency parameter $\beta$ that is owing to the uncertainty on the observational data. We also quantify the biases on the $\beta$ determination that is caused by the lack of constraints on the initial helium content and on the efficiencies of the superadiabatic convection and microscopic diffusion. 
}
{       
We adopted a modified grid-based SCEPtER pipeline to recover the $\beta$ parameter from synthetic stellar data.
Our grid spans the mass range [1.1; 1.6] $M_{\sun}$ and evolutionary 
stages from the zero-age main sequence (MS) to the central hydrogen depletion. The $\beta$ estimates were obtained by generalising the maximum
likelihood technique described in our previous works. As observational constraint, we adopted the effective temperatures, [Fe/H], masses, and radii of the two stars.
}
  {
By means of Monte Carlo simulations, adopting a reference scenario of mild overshooting $\beta = 0.2$ for the synthetic data, and taking typical observational errors  into account, we found both large statistical uncertainties and biases on the estimated values of $\beta$. For the first 80\% of the MS evolution, $\beta$ is biased by about $-0.04$, with the $1 \sigma$ error practically unconstrained in the whole explored range [0.0; 0.4]. In the last 5\% of the evolution the bias vanishes and the $1 \sigma$ error is about 0.05. 
The $1 \sigma$ errors are similar when adopting different reference values of $\beta$. Interestingly, for synthetic data computed without convective-core overshooting, the estimated $\beta$ is biased by about 0.12 in the first 80\% of the MS evolution, and by 0.05 afterwards. 
Assuming an uncertainty of $\pm 1$ in the helium-to-metal enrichment ratio $\Delta Y/\Delta Z$, we found a large systematic uncertainty in the recovered $\beta$ value, reaching 0.2 at the 60\% of the MS evolution.
Taking into account both the helium abundance indetermination  and $1 \sigma$ statistical uncertainty, we found that in the terminal part of the MS evolution the error on the estimated $\beta$ values ranges from $-0.05$ to $+0.10$, while $\beta$ is basically unconstrained throughout the explored range at earlier evolutionary stages. 
We quantified the impact of a uniform variation of $\pm 0.24$ in the mixing-length parameter $\alpha_{\rm ml}$ around the solar-calibrated value. The largest bias occurs in the last 5\% of the evolution with an error on the estimated median $\beta$ from $-0.03$ to $+0.07$. In this last part, the $1 \sigma$ uncertainty that addresses statistical and systematic error sources ranges from $-0.09$ to $+0.15$. 
Finally, we quantified the impact of a complete neglect of diffusion in the stellar evolution computations. In this case, the $1 \sigma$ uncertainty that addresses statistical and systematic error sources ranges from $-0.08$ to $+0.08$ in the terminal 5\% of the MS, while it is practically unconstrained in the first 80\% of the MS. 
 }
{The calibration of the convective core overshooting with double-lined eclipsing binaries - in the explored mass range and with both components still in their MS phase - appears to be poorly reliable, at least until further stellar observables, such as asteroseismic ones, and more accurate models are available.}


   \maketitle

\section{Introduction}\label{sec:intro}

As the result of the huge effort made in the past decades to refine the accuracy and reliability of the stellar evolutionary predictions, stellar evolution theory
 has become one of the most robust areas of astrophysical research.
However, several mechanisms involved in the evolution of stars are still poorly understood. 

The lack of a self-consistent treatment of convection in stellar evolutionary codes is one of the major weaknesses affecting stellar models. 
This lack prevents a firm and reliable prediction of the extension of the convective regions, 
both in the core and in the envelope. We focus on the convective-core extent during the central hydrogen burning phase of low-mass stars. 

In classical models the border of the convective core is determined by means of the Schwarzschild criterion. However, this border identifies the locus where the acceleration, 
not the velocity, of the convective element vanishes, thus raising the question of the extension of the overshoot beyond the border itself.  
The extension of  the extra-mixing region beyond the Schwarzschild border
is usually parametrized  in terms of the pressure scale height $H_{\rm 
        p}$: $l_{\rm ov} = \beta H_{\rm p}$, where $\beta$ is a free parameter. 

Many efforts to constrain the values of $\beta$ have been performed since the first investigations  \citep[among them we recall][]{Saslaw1965, Shaviv1973}. 
An obvious way to obtain clues on its value and on the possible dependence on the stellar mass is to compare 
the theoretical models computed with different convective-core overshooting efficiencies with proper observations.
Several methods for calibration have been explored in the literature. They include the isochrone fitting of stellar cluster colour-magnitude diagrams \citep[see e.g.][]{Bertelli1992, PradaMoroni2001, Barmina2002, Brocato2003, VandenBerg2006, Bressan2012} and the comparison with recently available asteroseismic constraints \citep[e.g.][]{Montalban2013, Guenther2014, Tkachenko2014, Aerts2015}.

Of the many possible candidates, double-line eclipsing binaries are often adopted as an ideal test bed for $\beta$ calibration. As an example, the relevance of the core overshooting efficiency was explored in \citet{Andersen1990}. By considering eight binary systems, they found that non-standard models were required for masses higher than 1.5 $M_{\sun}$. Other investigations, which tried to explore the dependence of $\beta$ on stellar mass, were performed by \citet{Ribas2000} and \citet{Claret2007}. These studies reached somewhat different conclusions. 
The investigation by \citet{Claret2007} indicates that the dependence of $\beta$ on mass is more uncertain and less steep than stated by \citet{Ribas2000}, reaching a plateau of $\beta \simeq 0.2$ for a star more massive than 2 $M_{\sun}$.
More recent investigations were performed by \citet{Meng2014} and \citet{Stancliffe2015}.
Overall, a low $\beta$ value (i.e. $ \lesssim $ 0.2) is typically considered enough to match the observational data.  

However, we recall that the $\beta$ value provided by the fitting procedure depends on the input physics actually implemented in the evolutionary code that is used 
to compute stellar models, even when the parametrization adopted to describe the overshooting is the same. Thus, it would be more meaningful to compare 
the convective-core extension instead of the $\beta$ values obtained by different sets of stellar models. 

Moreover, the aforementioned calibrations suffer from at least two shortcomings. The first, which makes a direct comparison of the results from different studies difficult,
is the lack of homogeneity in the treatment of the statistical errors owing to the stellar observational uncertainties. A chief difficulty is that the uncertainty in the derived masses is often neglected in the final error budget.  
The second problem is the lack of theoretical investigations on the possible bias in the value of $\beta$ obtained by a fit of the systems with stellar isochrones.

As a first effort to partially fill this gap, we here address many aspects of these questions. 
By means of Monte Carlo simulations, we provide some clues on the reliability of convective-core overshooting calibration for low-mass binary systems, restricting the analysis to the mass range [1.1; 1.6] $M_{\sun}$ and to the MS phase. This mass range contains several well-studied systems (e.g. AQ Ser, GX Gem, BK Peg, BW Aqr, V442 Cyg, AD Boo, VZ Hya, V570 Per, HD71636) that have been 
 used in literature to calibrate the $\beta$ parameter \citep[e.g.][]{Lacy2008,Clausen2010,Torres2014}.

We adopt the SCEPtER pipeline \citep{scepter1, eta, binary}, modified for estimating the core overshooting parameter. We address in a statistical robust way  the propagation of the uncertainty on the observational values to the fitted $\beta$ parameter and explore the possible biases that are due to some unconstrained mechanisms or input that influence the evolution in the considered mass range: the initial helium content of the stars, and the efficiency of microscopic diffusion and of the external convection, parametrized through the mixing-length value $\alpha_{\rm ml}$.

\section{Methods}\label{sec:method}

The value of the $\beta$ parameter was determined by means of the SCEPtER pipeline, a maximum-likelihood technique 
relying on a pre-computed grid of stellar models and on a set of observational constraints \citep[see e.g.][]{eta}. 
A first application to eclipsing binary systems has been extensively described in
\citet{binary}. We briefly summarize the technique here and focus on the modifications adopted to estimate $\beta$ .

We assume
${\cal S}_1$ and ${\cal S}_2$ to be two stars in a detached binary system. The observed quantities adopted to estimate the core overshooting efficiency are $q^{{\cal S}_{1,2}} \equiv \{T_{\rm eff, {\cal S}_{1,2}}, {\rm
        [Fe/H]}_{{\cal S}_{1,2}}, 
M_{{\cal S}_{1,2}}, R_{{\cal S}_{1,2}}\}$. Let $\sigma^{1,2} = \{\sigma(T_{\rm
        eff, {\cal S}_{1,2}}), \sigma({\rm [Fe/H]}_{{\cal S}_{1,2}}), \sigma(M_{{\cal S}_{1,2}}),
\sigma(R_{{\cal S}_{1,2}})\}$ be the uncertainty in the observed
quantities. For each point $j$ on the estimation grid of stellar models, 
we define $q^{j} \equiv \{T_{{\rm eff}, j}, {\rm [Fe/H]}_{j}, M_{j},
R_{j}\}$. 
Let $ {{\cal L}^{1,2}}_j$ be the single-star likelihood functions defined as
\begin{equation}
{{\cal L}^{1,2}}_j = \left( \prod_{i=1}^4 \frac{1}{\sqrt{2 \pi}
        \sigma^{1,2}_i} \right) 
\times \exp \left( -\frac{\chi_{1,2}^2}{2} \right)
\label{eq:lik}
,\end{equation}
where
\begin{equation}
\chi^2_{1,2} = \sum_{i=1}^4 \left( \frac{q_i^{{\cal S}_{1,2}} -
        q_i^j}{\sigma_i} \right)^2. 
\end{equation}

In the computation of the likelihood functions we consider only the grid points  
within $3 \sigma$ of all the variables from ${\cal S}_{1,2}$.
Moreover, in the estimation process we explicitly assume that the stars are coeval\footnote{This assumption is justified because in building the synthetic datasets we imposed that the age difference between the binary components must be $ \leq $ 10 Myr.}. 
We also impose that the two stars share a common value of $\beta$ since the sampling is performed at constant value of overshooting efficiency; in Sect.~\ref{sec:individual} we explore the effect of relaxing this constraint. Then, the joint likelihood  ${\cal \tilde L}$ of the system is computed as the product of the single star likelihood functions.
Let ${\cal \tilde L}_{\rm max}$ be the
maximum value obtained in this step. The joint-stars estimated $\beta$ is
obtained by averaging the corresponding quantities of all the models
with a likelihood greater than $0.95 \times {\cal \tilde L}_{\rm max}$.

In the computation  we assumed as standard
 deviations 100 K in $T_{\rm eff}$, 0.1 dex in [Fe/H], 1\% in mass, and 0.5\%
 in radius. The covariance matrix adopted in the perturbation accounts for a correlation of 0.95 between effective temperatures, 0.95 between metallicities, 0.8 between masses, and 0.0 between radii \citep[see][for details]{binary}.

\subsection{Stellar model grid}
The grid of models covers the evolution from the zero-age main-sequence (ZAMS) 
up to the central hydrogen depletion of stars with mass in the range [1.1; 1.6] $M_{\sun}$ and initial metallicity [Fe/H] from $-0.55$ dex to 0.55 dex. The grid was computed by means of the FRANEC stellar evolutionary code \citep{scilla2008, tognelli2011} in the same
configuration as was adopted to compute the Pisa Stellar
Evolution Data Base\footnote{\url{http://astro.df.unipi.it/stellar-models/}} 
for low-mass stars \citep{database2012, stellar}. 
Seventeen sets of models were computed with a varying core overshooting parameter $\beta$ from 0.0 to 0.4 (the last representing a possible
maximum value, see e.g. the results
in \citealt{Claret2007}) with a step of 0.025.
The details on the standard grid of stellar models, the sampling procedure, and the age estimation technique are fully described in 
\citet{eta} and \citet{binary}; the adopted input and the related uncertainties are discussed in \citet{cefeidi, incertezze1, incertezze2}.

\begin{figure*}
        \centering
        \includegraphics[height=18cm,angle=-90]{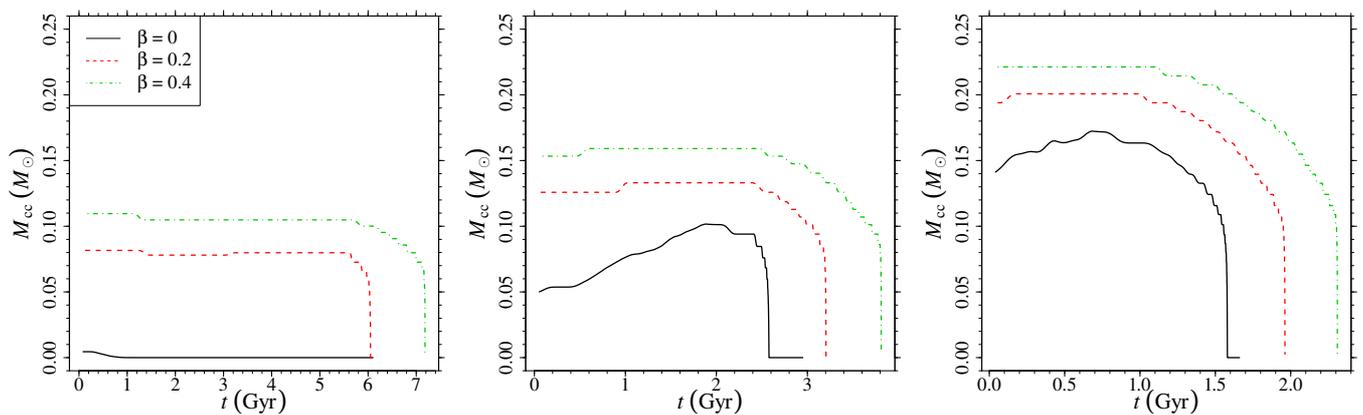}
        \caption{{\it Left}: extension of the convective core (in $M_ {\sun}$), from ZAMS to central hydrogen exhaustion, for three different values of the overshooting parameter $\beta$ = 0.0, 0.2, and 0.4 for a model of $M = 1.12$ $M_ {\sun}$ at initial [Fe/H] = 0.0. {\it Middle}: same as in the left panel, but for $M = 1.36$ $M_ {\sun}$. {\it Right}: same as in the left panel, but for $M = 1.60$ $M_ {\sun}$.}
        \label{fig:mcc}
\end{figure*}

\subsubsection{Influence of $\beta$ on the extension of the convective core}

Before we proceed with analysing the theoretical capability  of recovering an unknown value of $\beta$ from a synthetic dataset, it is worth showing the effect of the different choices of $\beta$ on the extension of the convective core for the models adopted in the present work. 
Figure~\ref{fig:mcc} shows for an initial [Fe/H] = 0.0 the evolution of the convective core mass versus the evolutionary time for stellar models of masses 1.12, 1.36, and 1.60 $M_{\sun}$, which covers the range we explore here. 
The extension of the convective core for the lightest model computed without overshooting is clearly negligible. Moreover, the impact of increasing $\beta$ on the maximum core extension is lower at higher masses. 
These results, coupled with the fact that more than 95\% of the {\it Kepler} targets in the KOI catalogue have a mass lower than about 1.6 $M_{\sun}$, are the main reasons for the choice of the upper and lower limits of the mass range explored here.

\section{Intrinsic accuracy and precision of the overshooting calibration}
\label{sec:res-random}

The first key point to discuss is the intrinsic precision and accuracy of the core overshooting calibration procedure, that is, in the ideal case in which the stellar models used in the recovery procedure perfectly match the data. In other words, it is important to establish 
the magnitude of both the random error and the possible bias in the recovered overshooting parameter that is due alone to the observational uncertainties that affect the observables used to constrain the overshooting itself. 
To this purpose, we followed a procedure similar to that outlined in \citet{binary}, as detailed below.

We determined the expected errors in the $\beta$ estimation process by Monte Carlo simulations. In the following we assume a true value of $\beta = 0.2$, regardless of the mass of the stars. This assumption would not affect the validity of the results if there were a trend of the overshooting parameter with the stellar mass \citep[e.g.][]{Ribas2000, Claret2007} because we are interested in a differential effect, that is, in the theoretical difference expected between the true and the recovered values of $\beta$.
We also discuss the effect of adopting different $\beta$ reference values in Sect.~\ref{sec:startpoints}.
We therefore generated a synthetic sample of $N = 50\,000$ binary systems from the grid of stellar models with $\beta = 0.2$ and subjected all these systems to random perturbation on the observables to simulate the effect of measurement errors. The overshooting parameter was then estimated from adopting the full grid of stellar models with seventeen different $\beta$ values in the range [0.0; 0.4]. It is important that that the procedure accounts simultaneously for all the errors in the observables, so that it is possible to obtain a robust estimate of $\beta$ and of its uncertainty.  

\begin{figure*}
\centering
\includegraphics[height=17cm,angle=-90]{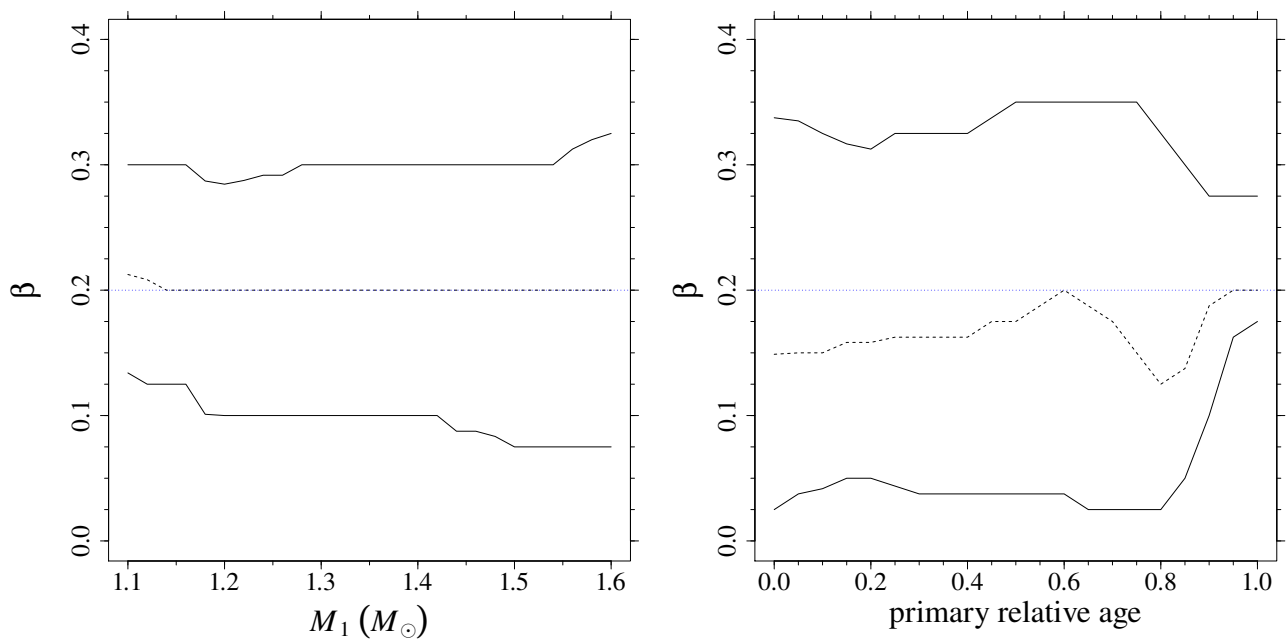}
\caption{{\it Left}: $1 \sigma$ envelope of $\beta$ estimates that are due to the observational errors (solid line) as a function of the mass of the primary star. The dashed line marks the position of the median. {\it Right}: same as in the \textup{\textup{{\it \textup{left}}} }panel, but as a function of the relative age of the primary star.
}
\label{fig:envelope-std}
\end{figure*}

\begin{figure*}
        \centering
        \includegraphics[height=17cm,angle=-90]{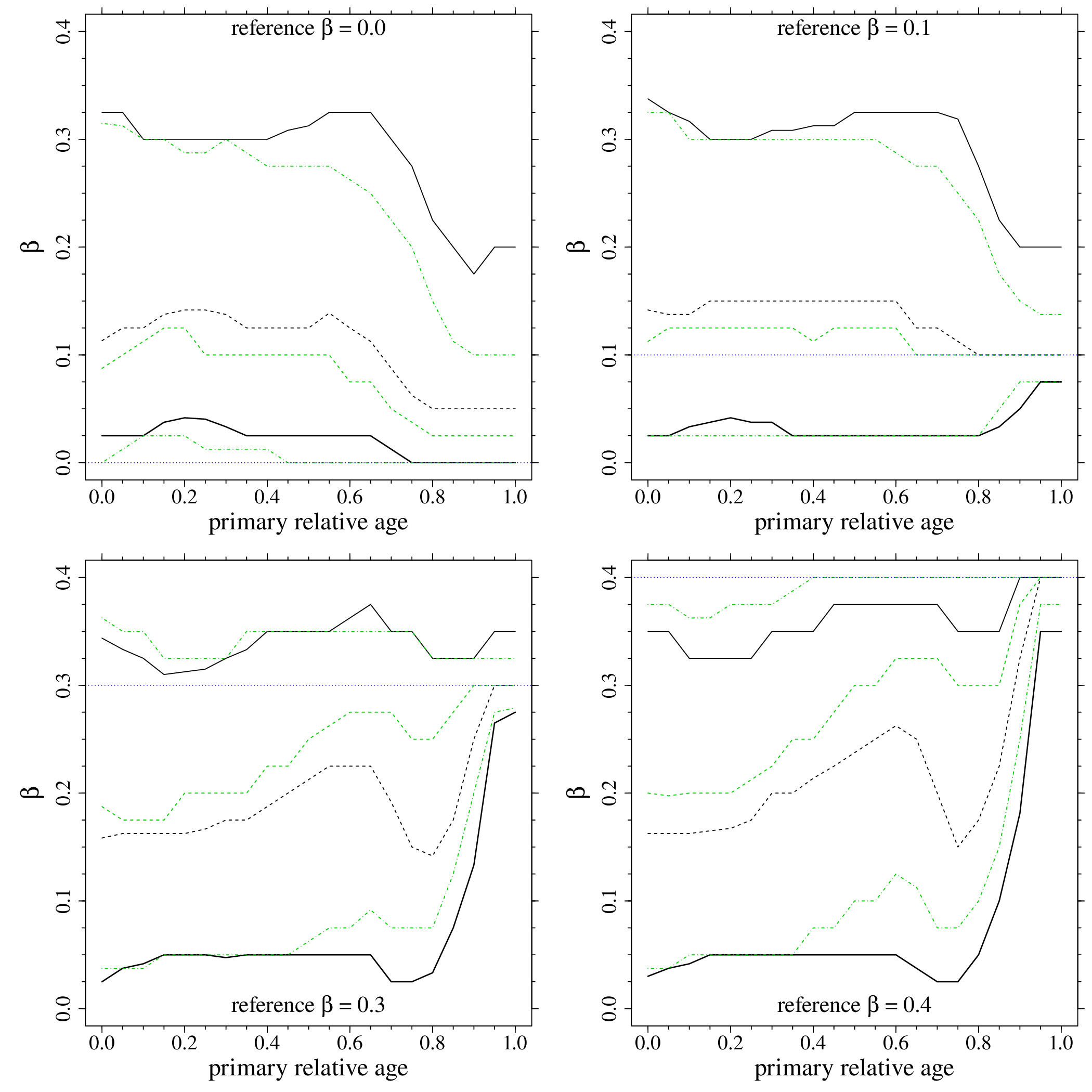}
        \caption{Same as in the right panel of Fig.~\ref{fig:envelope-std}, but sampling from the grid with different  $\beta$ values reported in the labels. The green dot-dashed lines refer to the scenario in which the nominal errors on the observables are halved. 
        }
        \label{fig:envelope-startpoints}
\end{figure*}

\begin{table*}[ht]
        \centering
        \caption{Median ($q_{50}$) and  $1 \sigma$ random envelope ($q_{16}$, $q_{84}$) of the estimated core overshooting parameter, as a function of the mass of the primary star, of its relative age $r$, and of the mass ratio of the system $q$. In all three cases, the envelope is obtained by marginalization over all the remaining parameters.}
        \label{tab:results}
        \begin{tabular}{lccccccccccc}
                \hline\hline
                \multicolumn{12}{c}{primary star mass ($M_{\sun}$)}\\
                & 1.1 & 1.2 & 1.3 & 1.4 & 1.5 & 1.6 & & & & & \\ 
                \hline
                $q_{16}$ & 0.13 & 0.10 & 0.10 & 0.10 & 0.07 & 0.07 & & & & & \\ 
                $q_{50}$ & 0.21 & 0.20 & 0.20 & 0.20 & 0.20 & 0.20 & & & & & \\ 
                $q_{84}$ & 0.30 & 0.28 & 0.30 & 0.30 & 0.30 & 0.32 & & & & & \\ 
                \hline
                \multicolumn{12}{c}{primary star relative age $r$}\\
                & 0.0 & 0.1 & 0.2 & 0.3 & 0.4 & 0.5 & 0.6 & 0.7 & 0.8 & 0.9 & 1.0 \\ 
                \hline
                $q_{16}$ & 0.03 & 0.04 & 0.05 & 0.04 & 0.04 & 0.04 & 0.04 & 0.03 & 0.03 & 0.10 & 0.17 \\ 
                $q_{50}$ & 0.15 & 0.15 & 0.16 & 0.16 & 0.16 & 0.17 & 0.20 & 0.17 & 0.12 & 0.19 & 0.20 \\ 
                $q_{84}$ & 0.34 & 0.33 & 0.31 & 0.33 & 0.33 & 0.35 & 0.35 & 0.35 & 0.33 & 0.28 & 0.28 \\ 
                \hline
                \multicolumn{12}{c}{mass ratio $q$}\\
                        & 0.65 & 0.70 & 0.75 & 0.80 & 0.85 & 0.90 & 0.95 & 1.0 & & & \\ 
                \hline
                $q_{16}$ & 0.07 & 0.07 & 0.06 & 0.06 & 0.06 & 0.06 & 0.10 & 0.11 & & & \\ 
                $q_{50}$ & 0.20 & 0.20 & 0.20 & 0.20 & 0.20 & 0.20 & 0.20 & 0.20 & & & \\ 
                $q_{84}$ & 0.35 & 0.33 & 0.32 & 0.31 & 0.31 & 0.30 & 0.30 & 0.30 & & & \\ 
                \hline
                & & & & & 
        \end{tabular}
\end{table*}

The results of the simulations are presented in Table~\ref{tab:results} and in Fig.~\ref{fig:envelope-std}. The table reports the median ($q_{50}$) of the estimated core overshooting parameter
and the $1 \sigma$ random error envelope ($q_{16}$ and $q_{84}$)
as a function of the mass of the primary star, of the primary star relative age $r$ (defined as the ratio between
the age of the star and the age of the same star at central hydrogen exhaustion), and of the mass ratio $q = M_2/M_1$. The envelope was obtained as in \citet{eta} by
computing the 16th and 84th quantiles of the 
relative errors over a moving window in mass and relative age. 

The main result is that the estimated $\beta$ values are affected by large random errors, with a mean envelope half-width of about 0.10.
We note a very mild increase of the envelope width as a function of the mass of the primary star; with respect to the mass ratio $q,$ we note a mild shrinking of the envelope at high $q$. 
The most significant trend is due to the evolutionary stage of the primary star. While the overshooting parameter is consistently underestimated by about 0.04 when the primary star is in the early part of its MS phase ($r \leqq 0.5$), in a more advanced MS stage ($r \gtrsim 0.7$) we observe a trend similar to that described in \citet{eta} for mass estimates. 
In this zone the morphology of the grids computed with various overshooting efficiencies are most different, since the overall contraction starts in different regions of the grids.
As a consequence, it is more likely that a model computed with overshooting efficiency $\beta$ is nearer to a model with $\tilde{\beta} < \beta$ than to a model with  $\tilde{\beta} > \beta$. This effect leads to the observed underestimation of $\beta$.  In the very last evolutionary stages ($r \geq 0.9$) the bias vanishes since all the stellar tracks return to evolve parallel after the overall contraction end. In this zone the error envelope half-width shrinks from a mean value of 0.15 to about 0.06.

The apparent lack of bias in the left panel of Fig.~\ref{fig:envelope-std} should be considered with caution. It is due to the adopted sampling method, which favours systems with primary stars in the later evolutionary stages, where the bias in $\beta$ is absent \citep[see][for a detailed discussion on the adopted sampling]{binary}. A different sampling would show the biases present at low $r$.

In summary, even in this ideal case of a perfect agreement between the stellar tracks adopted for the recovery procedure and the artificial stars, which are sampled from the same grid of models, the estimated value of $\beta$ is generally biased towards lower values and shows a large uncertainty. 

Such a considerable random error undermines the possibility of obtaining a reliable estimate of the core overshooting parameter from a single binary system in the considered mass range with stars in the MS because the probability of calibrating on a statistical fluctuation is high. 
Moreover, the presence of the bias sheds doubts even on the statistical calibration of $\beta$ that is performed by combining the results obtained in several systems.

\subsection{Effect of changing the $\beta$ reference value}
\label{sec:startpoints}

In this section we briefly present some results obtained by adopting different  $\beta$ reference values and discuss the differences with respect to the standard scenario of $\beta = 0.2$. Figure~\ref{fig:envelope-startpoints} -- analogous to the right panels of Fig.~\ref{fig:envelope-std} -- shows the resulting error envelope obtained by assuming  $\beta$ = 0.0, 0.1, 0.3, and 0.4 as
reference values. The differences between the panel are apparent. 

In the case of sampling from the grid with $\beta = 0.0$, no underestimation can occur; therefore the results are influenced by an edge effect similar to those discussed in other works for mass, radius, and age estimates \citep{scepter1,eta,binary}.
A median bias of about 0.12 was found up to $r \approx 0.8$. In the last part of the evolution the bias is of about 0.05. For comparison we also plot in the figure the envelope obtained by halving the uncertainties on the observational parameters. In this case the bias is lower and the envelope shrinks, but only for $r \geq 0.4$.
Concerning the effect of the single error sources, we verified that their individual halving contributes in nearly equal manner to the final result.

This scenario is particularly interesting because it shows the tendency to prefer models with moderate overshooting efficiency even in the fit of binary systems sampled from a grid without overshooting.  The resulting median bias 
is relevant since in the recent literature, as discussed in Sect.~\ref{sec:intro}, there is a general feeling that $\beta \leq 0.2$ is adequate to describe the observations.

The presence of a distortion in the estimates of the convective-core overshooting parameters sheds some doubts on the precision of the estimates at low $\beta$, in particular for $\beta \leq 0.1$.

The case of reference $\beta = 0.1$ provides a lower biases than all other scenarios in the estimated overshooting efficiency; the true value is overestimated by about 0.05 up to $r \approx 0.8$. However, the estimate is affected by non-negligible random errors; even in the last part of the evolution -- where the precision is highest -- the $1 \sigma$ error envelope is wide (from overestimation of $+0.1$ to underestimation of $-0.025$).  

The cases of reference $\beta = 0.3, 0.4$ are more similar to the standard scenario, showing a strong underestimation (greater than 0.2 for $\beta = 0.4$) in the first part of the evolution, and the characteristic underestimation around $r = 0.8$. In both cases the latter underestimation is partially suppressed when the errors on the observables are halved.

In conclusion, while the expected bias depends on the actual value of the overshooting efficiency in the reference set, there is clear evidence that all the scenarios show a large variability in the $\beta$ estimates at $1 \sigma$ level. Furthermore, the bias toward higher values of $\beta$ that we found when sampling from $\beta = 0.0$ suggests great care in the evaluation of fitted $\beta \lesssim 0.1$.

\subsection{Effect of relaxing the assumption of a common $\beta$ for both stars}
\label{sec:individual}

In the previous analyses we assumed that the two stellar components share a common value of $\beta$. Since it has been proposed in the literature that a possible trend of $\beta$ with the stellar mass exists \citep{Ribas2000, Claret2007}, we verify in this section that no biases mar our results that are caused by the estimation procedure assumptions.

To this purpose we repeated the estimation of the overshooting parameter for $N = 10\,000$ binary systems, sampled from the grid of $\beta = 0.2$,  without imposing the constraint that the solutions for the primary and the secondary stars share a common $\beta$ value. In other words, only the coevality was imposed during the recovery. This procedure would show possible systematic distortions in the estimated overshooting parameter for the two stars in the systems. A lack of bias would support the conclusion that the individual $\beta$ is recovered well by our technique. 

\begin{figure*}
        \centering
        \includegraphics[height=17cm,angle=-90]{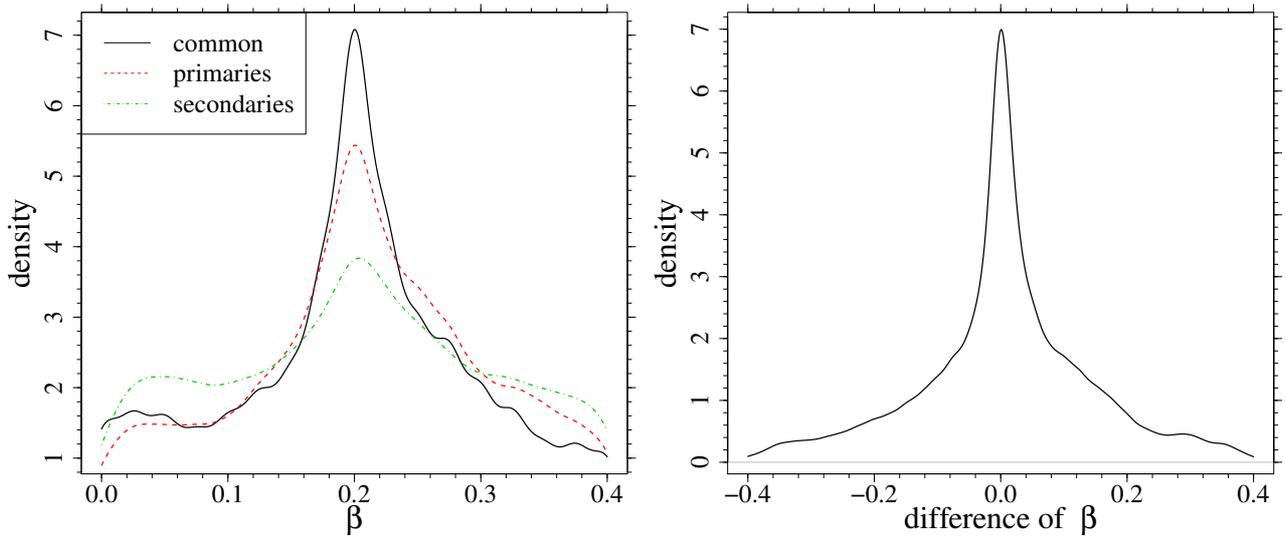}
        \caption{{\it Left}: probability density functions for the estimated $\beta$ from $N$ = 10\,000 systems sampled at true $\beta = 0.2$. The black solid line corresponds to the estimation performed assuming a common value of $\beta$ for primary and secondary star; the red dashed line corresponds to the $\beta$ values for the primary star alone; the greed dot-dashed line to  $\beta$ values for the secondary star alone. {\it Right}: probability density function for the differences of the estimated $\beta$ for primary and secondary stars.
        }
        \label{fig:density-individual}
\end{figure*}

The results of the analysis are displayed in Fig.~\ref{fig:density-individual}. The left panel of the figure shows the probability density function (evaluated by means of a kernel density estimator, see e.g. \citealt{Scott1992, 
venables2002modern} and Appendix A in \citealt{scepter1}) of the recovered $\beta$ values for the whole systems -- that is, imposing that the two stars share a common value, as in the previous sections -- and individually for primary and secondary stars. It is apparent that the recovered values are always consistent in median with the true value $\beta = 0.2$ and that the distributions of the individual estimates are wider than the common one, especially for the secondary stars, which are sampled in earlier stages of the their evolution.  The right panel of Fig.~\ref{fig:density-individual} shows the probability density function of the $\beta$ differences between primary and secondary stars for the $N $ binary systems. The function is strongly peaked at zero difference, with the 16th and 84th quantile $\mp 0.12$ respectively.

These results are reassuring for the capability of the technique of recovering the correct value of the overshooting parameter for the two stars, without suffering from distortions that are
due to the different masses and evolutionary stages from the primary and secondary star.     

In the light of these results, it is safe to conclude that the individual $\beta$ for the two stars can be estimated without significant bias with respect to the scenario of constrained common value, at the expense of an extra loss of precision on the individual values. 

However, since no definitive trend of $\beta$ with the stellar mass is firmly established, we prefer to adopt in the sampling scheme that we followed to produce the synthetic dataset a common $\beta$ value for both components and impose the same constraint in the recovery phase as well. In turn, this choice translates into a conservative
evaluation of the uncertainties in the estimated $\beta$ values.

\section{Stellar models induced biases in the calibration procedure}
\label{sec:bias}

In addition to the random fluctuation discussed in the previous section, the estimated $\beta$ value is prone to systematic biases that are due to stellar models adopted in the calibration procedure. Several inputs or assumptions are routinely adopted in stellar evolutionary codes, which are still poorly constrained by observations. 
In this section we focus on three bias sources that are relevant in the mass range considered in this paper: the initial helium content, the mixing-length value, and the efficiency of the element diffusion used to compute stellar models. While the influence of the last two is expected to disappear for more massive systems
because the convective envelope vanishes and because the timescale evolves very fast, the first is unavoidable in every mass range.    

To quantify the biases due to the considered factors, we followed the same procedure as described in \citet{binary}. We built several synthetic datasets of artificial binary systems sampling from grids computed with $\beta = 0.2$, but varying the initial helium content, the mixing-length value, or neglecting microscopic diffusion. These systems were subjected to random perturbation to simulate the observational errors, following the same procedure as outlined in Sect.~\ref{sec:res-random}.
Finally, we estimated the overshooting parameter by adopting the standard multi-grid. 

\subsection{Initial helium abundance}
\label{sec:He}

The evolution of stars strongly depends on the initial helium abundance. As a consequence, for a given mass 
and age, the observable stellar quantities depend on its value. On the other hand, the stars we examine here are too cold to allow a spectroscopic measurement of the surface helium content. Moreover, such a value 
would not be representative of the initial value owing to the effect of microscopic diffusion. 
 
This means that stellar modellers have to assume an initial value
$Y$ for the helium abundance to adopt in stellar evolution computations. A linear relationship between the
primordial helium abundance $Y$ and metallicity $Z$ is usually
assumed. Unfortunately, the value of the helium-to-metal enrichment ratio $\Delta Y/\Delta Z$ is still
quite uncertain \citep[e.g.][]{pagel98,jimenez03,gennaro10}. This in turn leads to
an uncertainty in the initial helium abundance adopted in stellar computations
for a given initial metallicity, which affects stellar model predictions and, consequently, the value of 
the estimated $\beta$ parameter.

To assess the bias that is due to the uncertain initial helium content, we computed two additional grids of stellar models with $\beta = 0.2$,
the same metallicity values $Z$ as in the standard grid, but by
changing the helium-to-metal enrichment ratio $\Delta Y/\Delta Z$ to values 1
and 3. Then, we built two synthetic datasets, each of $N = 50\,000$ artificial
binary systems by sampling the objects from these two non-standard grids, and subjected the observables of the obtained systems to random perturbations. 
As a last step, the core overshooting efficiency was reconstructed using the standard multi-grid with $\Delta Y/\Delta Z = 2$.   

\begin{figure*}
        \centering
        \includegraphics[height=17cm,angle=-90]{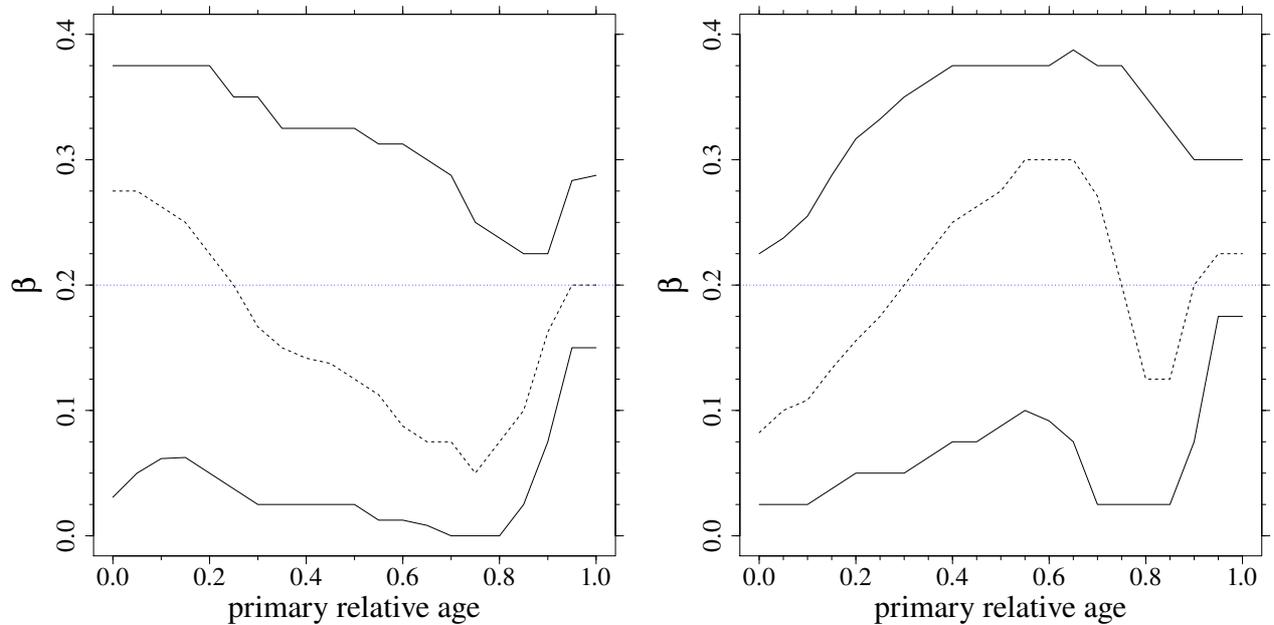}
        \caption{{\it Left}: as in the right panel of Fig.~\ref{fig:envelope-std}, but sampling from a grid with $\Delta Y/\Delta Z = 1$. {\it Right}: same as in the {\it left} panel, but sampling from a grid with $\Delta Y/\Delta Z = 3$. In both cases the reconstruction was performed with the standard grid with $\Delta Y/\Delta Z = 2$.
        }
        \label{fig:envelope-He}
\end{figure*}

\begin{figure*}
        \centering
        \includegraphics[height=19cm,angle=-90]{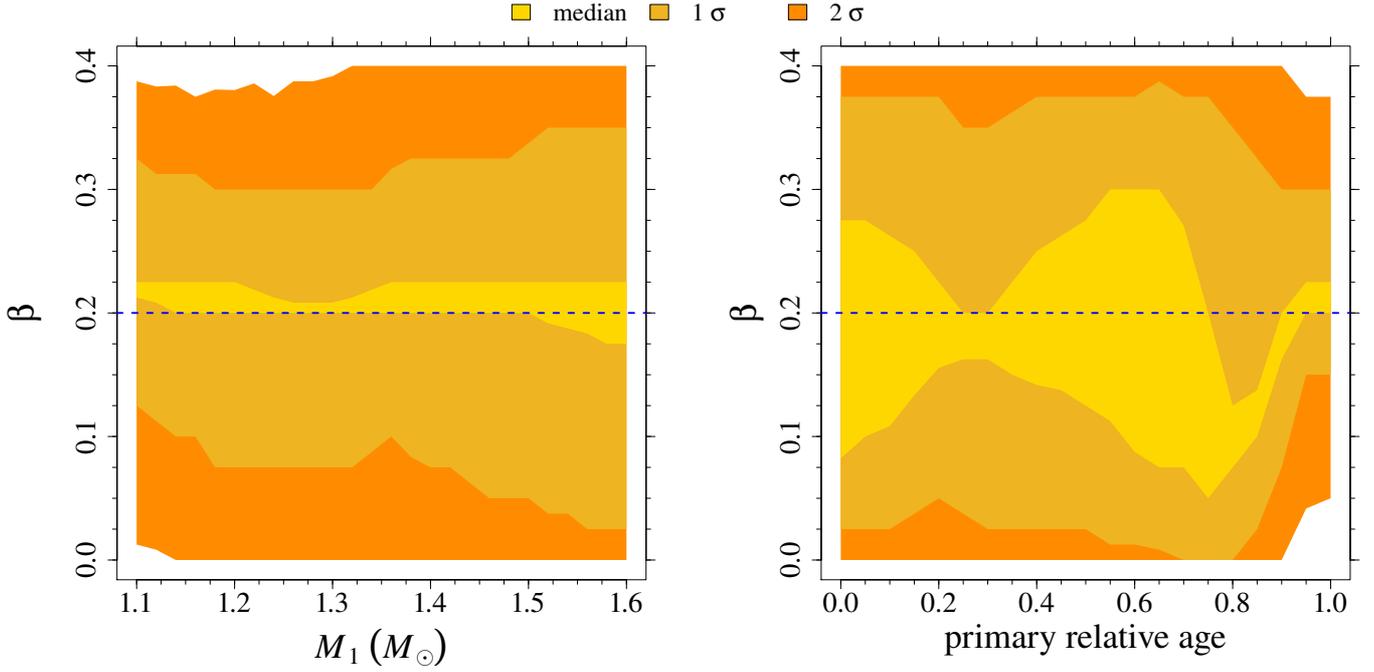}
        \caption{{\it Left}: overall uncertainty (statistical and systematic) that is due to the considered variations in $\Delta Y/\Delta Z$, in dependence on the mass of the primary star. The lighter region shows the uncertainty on the median of the estimated $\beta$ caused by ignoring the correct initial helium value. The intermediate colour regions show the overall error up to $1 \sigma$  that is due to the cumulated contribution of observational errors and systematic bias as a result of the unknown initial helium value. The darker regions correspond to cumulated errors up to $2 \sigma$.  {\it Right}: same as in the {\it left} panel, but in dependence on the relative age of the primary star.
        }
        \label{fig:envelope-He-col}
\end{figure*}

\begin{table}[ht]
        \centering
        \caption{Overall uncertainty, statistical ($1 \sigma$ and $2 \sigma$) and systematic ($q_{50}^{l}$, $q_{50}^{u}$) as
a result of the considered variations in $\Delta Y/\Delta Z$, $\alpha_{\rm ml}$, and microscopic diffusion efficiency as a function of the mass of the primary star.} 
        \label{tab:bias-he-ml-M}
        \begin{tabular}{rcccccc}
                \hline\hline
                \multicolumn{7}{c}{primary star mass ($M_{\sun}$)}\\
                & 1.1 & 1.2 & 1.3 & 1.4 & 1.5 & 1.6 \\ 
                \hline
                \multicolumn{7}{c}{unknown initial helium content}\\
                $q_{0.025}$ & 0.01 & 0.00 & 0.00 & 0.00 & 0.00 & 0.00 \\ 
                $q_{0.16}$ & 0.12 & 0.07 & 0.07 & 0.07 & 0.05 & 0.03 \\ 
                $q_{50}^{l}$ & 0.21 & 0.20 & 0.20 & 0.20 & 0.20 & 0.17 \\ 
                $q_{50}^{u}$ & 0.23 & 0.22 & 0.21 & 0.23 & 0.23 & 0.23 \\ 
                $q_{0.84}$ & 0.32 & 0.30 & 0.30 & 0.33 & 0.34 & 0.35 \\ 
                $q_{0.975}$ & 0.39 & 0.38 & 0.39 & 0.40 & 0.40 & 0.40 \\ 
                \hline
                \multicolumn{7}{c}{unknown mixing-length}\\
                $q_{0.025}$ & 0.01 & 0.00 & 0.00 & 0.00 & 0.00 & 0.00 \\ 
                $q_{0.16}$ & 0.09 & 0.07 & 0.07 & 0.07 & 0.07 & 0.07 \\ 
                $q_{50}^{l}$ & 0.17 & 0.16 & 0.17 & 0.17 & 0.18 & 0.20 \\ 
                $q_{50}^{u}$ & 0.30 & 0.28 & 0.25 & 0.23 & 0.23 & 0.20 \\ 
                $q_{0.84}$ & 0.38 & 0.36 & 0.35 & 0.33 & 0.33 & 0.33 \\ 
                $q_{0.975}$ & 0.40 & 0.40 & 0.40 & 0.40 & 0.40 & 0.40 \\ 
                \hline
                \multicolumn{7}{c}{unknown element diffusion efficiency}\\
                $q_{0.025}$ & 0.01 & 0.00 & 0.00 & 0.00 & 0.00 & 0.00 \\ 
                $q_{0.16}$ & 0.10 & 0.05 & 0.05 & 0.07 & 0.07 & 0.07 \\ 
                $q_{50}^{l}$ & 0.17 & 0.17 & 0.17 & 0.17 & 0.18 & 0.20 \\ 
                $q_{50}^{u}$ & 0.21 & 0.20 & 0.20 & 0.20 & 0.20 & 0.20 \\ 
                $q_{0.84}$ & 0.30 & 0.28 & 0.30 & 0.30 & 0.32 & 0.33 \\ 
                $q_{0.975}$ & 0.38 & 0.38 & 0.38 & 0.40 & 0.40 & 0.40 \\ 
                \hline
        \end{tabular}
\end{table}

\begin{table*}[ht]
        \centering
        \caption{Same as Table \ref{tab:bias-he-ml-M}, but as a function of the relative age of the primary star.} 
        \label{tab:bias-he-ml-r}
        \begin{tabular}{rccccccccccc}
                \hline\hline
                \multicolumn{12}{c}{primary star relative age}\\
                & 0.0 & 0.1 & 0.2 & 0.3 & 0.4 & 0.5 & 0.6 & 0.7 & 0.8 & 0.9 & 1.0 \\ 
                \hline
                \multicolumn{12}{c}{unknown initial helium content}\\
                $q_{0.025}$ & 0.00 & 0.00 & 0.00 & 0.00 & 0.00 & 0.00 & 0.00 & 0.00 & 0.00 & 0.00 & 0.05 \\ 
                $q_{0.16}$ & 0.03 & 0.03 & 0.05 & 0.03 & 0.03 & 0.03 & 0.01 & 0.00 & 0.00 & 0.07 & 0.15 \\ 
                $q_{50}^{l}$ & 0.08 & 0.11 & 0.16 & 0.16 & 0.14 & 0.12 & 0.09 & 0.07 & 0.07 & 0.16 & 0.20 \\ 
                $q_{50}^{u}$ & 0.28 & 0.26 & 0.23 & 0.20 & 0.25 & 0.28 & 0.30 & 0.27 & 0.12 & 0.20 & 0.23 \\ 
                $q_{0.84}$ & 0.38 & 0.38 & 0.38 & 0.35 & 0.38 & 0.38 & 0.38 & 0.38 & 0.35 & 0.30 & 0.30 \\ 
                $q_{0.975}$ & 0.40 & 0.40 & 0.40 & 0.40 & 0.40 & 0.40 & 0.40 & 0.40 & 0.40 & 0.40 & 0.38 \\ 
                \hline
        \multicolumn{12}{c}{unknown mixing-length}\\
                $q_{0.025}$ & 0.00 & 0.00 & 0.00 & 0.00 & 0.00 & 0.00 & 0.00 & 0.00 & 0.00 & 0.00 & 0.03 \\ 
                $q_{0.16}$ & 0.01 & 0.03 & 0.04 & 0.03 & 0.03 & 0.04 & 0.03 & 0.03 & 0.03 & 0.07 & 0.11 \\ 
                $q_{50}^{l}$ & 0.12 & 0.12 & 0.14 & 0.13 & 0.12 & 0.17 & 0.19 & 0.15 & 0.12 & 0.15 & 0.17 \\ 
                $q_{50}^{u}$ & 0.15 & 0.17 & 0.19 & 0.19 & 0.20 & 0.23 & 0.20 & 0.17 & 0.14 & 0.23 & 0.27 \\ 
                $q_{0.84}$ & 0.35 & 0.35 & 0.35 & 0.34 & 0.35 & 0.36 & 0.35 & 0.35 & 0.34 & 0.30 & 0.35 \\ 
                $q_{0.975}$ & 0.40 & 0.40 & 0.40 & 0.40 & 0.40 & 0.40 & 0.40 & 0.40 & 0.40 & 0.40 & 0.40 \\ 
                \hline
                \multicolumn{12}{c}{unknown element diffusion efficiency}\\
                $q_{0.025}$ & 0.00 & 0.00 & 0.00 & 0.00 & 0.00 & 0.00 & 0.00 & 0.00 & 0.00 & 0.00 & 0.01 \\ 
                $q_{0.16}$ & 0.03 & 0.03 & 0.04 & 0.04 & 0.04 & 0.04 & 0.04 & 0.03 & 0.03 & 0.05 & 0.12 \\ 
                $q_{50}^{l}$ & 0.13 & 0.13 & 0.15 & 0.16 & 0.16 & 0.17 & 0.20 & 0.17 & 0.12 & 0.17 & 0.17 \\ 
                $q_{50}^{u}$ & 0.15 & 0.15 & 0.16 & 0.17 & 0.19 & 0.23 & 0.28 & 0.26 & 0.20 & 0.19 & 0.20 \\ 
                $q_{0.84}$ & 0.34 & 0.33 & 0.31 & 0.33 & 0.35 & 0.36 & 0.38 & 0.38 & 0.36 & 0.31 & 0.28 \\ 
                $q_{0.975}$ & 0.40 & 0.40 & 0.40 & 0.40 & 0.40 & 0.40 & 0.40 & 0.40 & 0.40 & 0.40 & 0.38 \\ 
                \hline
        \end{tabular}
\end{table*}

The resulting error envelopes are presented in Fig.~\ref{fig:envelope-He} as a function of the relative age of the primary star. The median estimated $\beta$ in the low-helium scenario tends to decrease with $r$ up to $r \simeq 0.7$, while the opposite trend appears for the high initial helium case. We also note a shift in the estimated $\beta$ for stars near to the ZAMS, positive for the $\Delta Y/\Delta Z = 1$ case, and negative for $\Delta Y/\Delta Z = 3$. The trends visible in the figure are obviously caused by the differential effects on the stellar evolution timescale of changing the initial helium content and of assuming a different overshooting efficiency $\beta$.
The great difference between the two panels of Fig.~\ref{fig:envelope-He} poses a serious problem whenever the $\beta$ value is attempted
to be constrained because the original helium content of a system is generally poorly constrained. 

The final uncertainty on $\beta$ for a change of $\pm 1$ in $\Delta Y/\Delta Z$  is presented in Fig.\ref{fig:envelope-He-col} and Tables~\ref{tab:bias-he-ml-M} and \ref{tab:bias-he-ml-r}. The figure and tables show the position of the median and of the $1 \sigma$ and $2 \sigma$ error boundaries on the $\beta$ estimates, accounting for the chosen initial helium variability. They were constructed by computing for each relative age $r$ the maximum and minimum values assumed by the envelope boundaries and by the median in all the considered initial helium scenarios.
As an example, the $q_{50}^{l}$ ($q_{50}^{u}$) value was obtained by considering the minimum (maximum) value at a given mass (Table~\ref{tab:bias-he-ml-M}) or $r$ (Table~\ref{tab:bias-he-ml-r}) of the median $\beta$ obtained sampling from $\Delta Y/\Delta Z$ = 1, 2, 3 grids.

We note a mild increase of the uncertainty with the mass of the primary star $M_1$; the trend with $r$ is clearly more problematic and suggests that for relative ages $r \leq 0.8$ the effect of the unknown initial helium content even prevents the possibility of determining the sign of the bias. For the very last evolutionary phases the $1 \sigma$ random error on $\beta$ ranges from $-0.05$ to $+0.10$, while at the $2 \sigma$ level the value of $\beta$ is almost unconstrained in the whole explored range of overshooting efficiency.

\subsection{Mixing-length value}
\label{sec:ML}
One of the weakest point in stellar model computations is the treatment of the convective transport
in superadiabatic regimes, which actually prevents a firm
prediction of the effective temperature and radius of stars with an outer convective
envelope. 

Stellar evolutionary codes usually implement the mixing-length formalism \citep{bohmvitense58}, where the
efficiency of the convective transport depends on a free parameter
$\alpha_{\rm ml}$ that is routinely calibrated on the Sun,  a
procedure that in our standard case provides $\alpha_{\rm
  ml} = 1.74$. Nevertheless, there are several doubts
   that the solar-calibrated $\alpha_{\rm ml}$ value is also suitable
for stars of different masses and/or in different evolutionary stages \citep[see e.g.][]{Ludwig1999,Trampedach2014}.

To establish the impact of assuming different values of  superadiabatic convective efficiency, 
we computed two additional grids of stellar models with $\beta = 0.2$, but with $\alpha_{\rm ml} = 1.50$ and $\alpha_{\rm ml} = 1.98$. 
As in the previous section, we built two synthetic datasets, each of $N = 50\,000$ artificial binary systems, 
by sampling the objects from these two non-standard grids, and we subjected the observables to random perturbations. 
Finally, the overshooting efficiency was reconstructed using the standard multi-grid with $\alpha_{\rm ml} = 1.74$.  

The assumed scenario is appropriate for systems with $q \approx 1$; these stars are expected to share a common value -- whatever it is -- of mixing-length. The hypothesis is more questionable for systems with significantly different masses, since a possible trend of $\alpha_{\rm ml}$  with the stellar mass is still uncertain \citep{Trampedach2011, Bonaca2012, Mathur2012,Tanner2014,Magic2014}. Therefore the results presented here can be considered as the extreme variation on the $\beta$ calibration for differences in the mixing-length parameter of $\pm 0.24$.  

\begin{figure*}
        \centering
        \includegraphics[height=17cm,angle=-90]{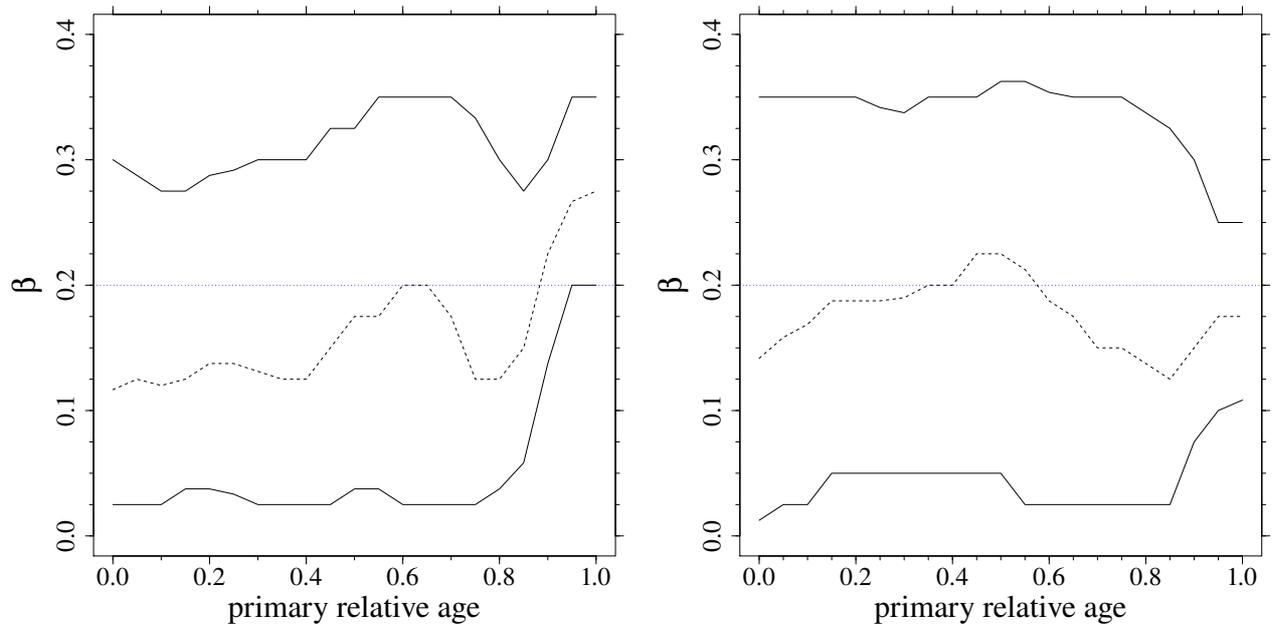}
        \caption{{\it Left}: as in the right panel of Fig.~\ref{fig:envelope-std}, but sampling from a grid with $\alpha_{\rm ml} = 1.50$. {\it Right}: same as in the {\it left} panel, but sampling from a grid with $\alpha_{\rm ml} = 1.98$. In both cases the reconstruction was performed with the standard grid with $\alpha_{\rm ml} = 1.74$.
        }
        \label{fig:envelope-ML}
\end{figure*}

\begin{figure*}
        \centering
        \includegraphics[height=19cm,angle=-90]{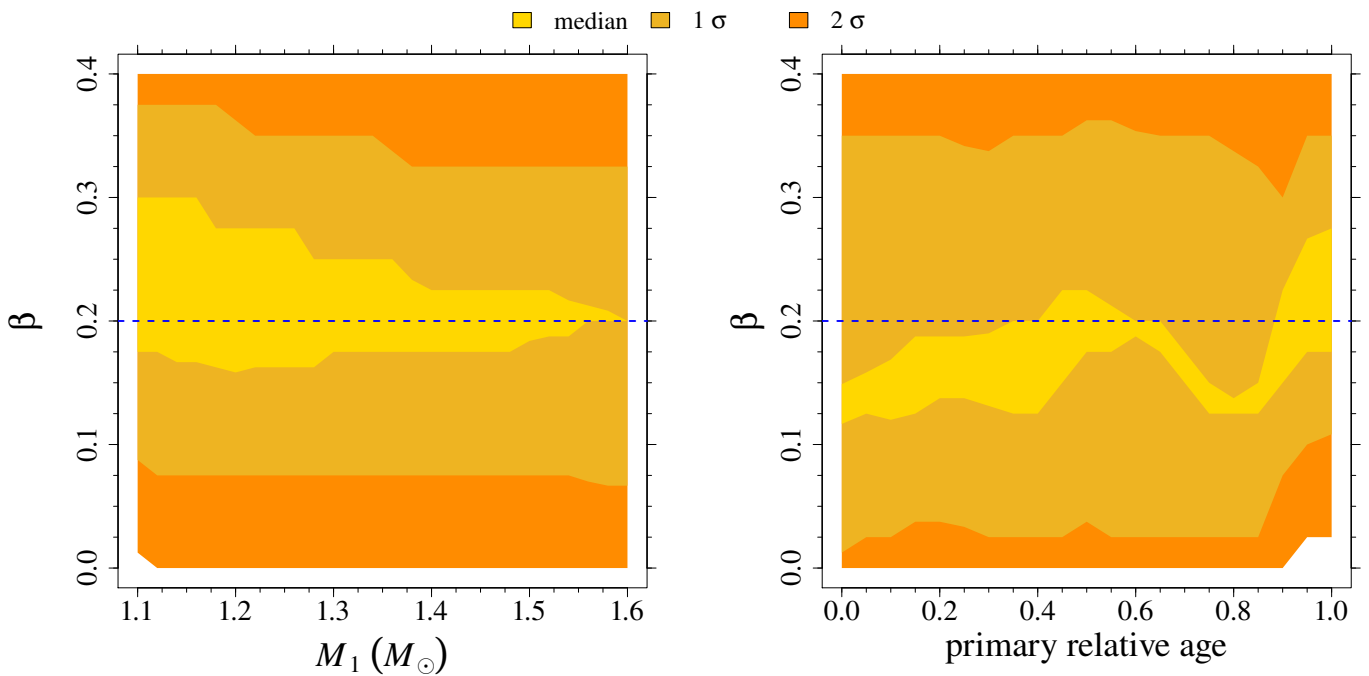}
        \caption{{\it Left}: overall uncertainty (statistical and systematic) that is due to the considered variations in $\alpha_{\rm ml}$, in dependence on the mass of the primary star. The lighter region shows the uncertainty on the median of the estimated $\beta$ that is a result of ignoring the correct mixing-length value. The intermediate colour regions show the overall error up to $1 \sigma$  caused by the cumulated contribution of observational errors and systematic bias that is due to the unknown mixing-length value. The darker regions correspond to errors up to $2 \sigma$.  {\it Right}: same as in the {\it left} panel, but in dependence on the relative age of the primary star.
        }
        \label{fig:envelope-ML-col}
\end{figure*}

The error envelopes as a function of the relative age of the primary star are shown in Fig.~\ref{fig:envelope-ML}. The trends caused by assuming different mixing-length values are slightly different; lowering $\alpha_{\rm ml}$ has the effect of underestimating the $\beta$ value for $r \leq 0.7$, while for $r \geq 0.9$ the overshooting efficiency is overestimated. In contrast, an higher value of $\alpha_{\rm ml}$ leads to a $\beta$ underestimate in the last part of the MS evolution.  
 
The final uncertainty on $\beta$ that is due to the ignorance of $\alpha_{\rm ml}$ is presented in Fig. \ref{fig:envelope-ML-col} and Tables~\ref{tab:bias-he-ml-M} and \ref{tab:bias-he-ml-r}.
As expected (see left panel in Fig.~\ref{fig:envelope-ML-col}), the bias caused by the explored uncertainty in the mixing-length value vanishes for massive systems. In these cases the primary stars do not present a convective envelope. 

Finally (right panel in Fig.~\ref{fig:envelope-ML-col}), a moderate bias of about $\pm 0.05$ in the last stages of the evolution is apparent, an unfortunate event, since this zone has been found in the preceding sections to offer the best opportunity for the $\beta$ estimate. 

\begin{table*}[ht]
        \centering
        \caption{Same as Table \ref{tab:bias-he-ml-r}, but only for the impact of $\alpha_{\rm ml}$ and diffusion efficiency, restricting the mass range of the secondary stars to $M_2 > 1.4$ $M_{\sun}$ and $M_2 > 1.5$ $M_{\sun}$.} 
        \label{tab:bias-ml-M2}
        \begin{tabular}{rccccccccccc}
                \hline\hline
                \multicolumn{12}{c}{primary star relative age}\\
                & 0.0 & 0.1 & 0.2 & 0.3 & 0.4 & 0.5 & 0.6 & 0.7 & 0.8 & 0.9 & 1.0 \\ 
                \hline
                \multicolumn{12}{c}{$M_2 > 1.4$ $M_{\sun}$}\\
                \hline
                \multicolumn{12}{c}{unknown mixing-length value}\\
                $q_{0.025}$ & 0.00 & 0.00 & 0.00 & 0.00 & 0.00 & 0.00 & 0.00 & 0.00 & 0.00 & 0.00 & 0.07 \\ 
                $q_{0.16}$ & 0.03 & 0.04 & 0.04 & 0.03 & 0.03 & 0.03 & 0.03 & 0.03 & 0.03 & 0.07 & 0.15 \\ 
                $q_{50}^{l}$ & 0.12 & 0.15 & 0.15 & 0.12 & 0.14 & 0.17 & 0.19 & 0.17 & 0.15 & 0.17 & 0.19 \\ 
                $q_{50}^{u}$ & 0.15 & 0.17 & 0.16 & 0.17 & 0.17 & 0.20 & 0.19 & 0.20 & 0.17 & 0.23 & 0.24 \\ 
                $q_{0.84}$ & 0.33 & 0.32 & 0.30 & 0.34 & 0.35 & 0.38 & 0.38 & 0.36 & 0.38 & 0.34 & 0.30 \\ 
                $q_{0.975}$ & 0.40 & 0.40 & 0.40 & 0.40 & 0.40 & 0.40 & 0.40 & 0.40 & 0.40 & 0.40 & 0.40 \\ 
                \hline
                \multicolumn{12}{c}{unknown element diffusion efficiency}\\
                $q_{0.025}$ & 0.00 & 0.00 & 0.00 & 0.00 & 0.00 & 0.00 & 0.00 & 0.00 & 0.00 & 0.00 & 0.09 \\ 
                $q_{0.16}$ & 0.04 & 0.05 & 0.04 & 0.03 & 0.03 & 0.03 & 0.03 & 0.03 & 0.03 & 0.07 & 0.15 \\ 
                $q_{50}^{l}$ & 0.15 & 0.15 & 0.16 & 0.17 & 0.17 & 0.17 & 0.19 & 0.20 & 0.17 & 0.20 & 0.17 \\ 
                $q_{50}^{u}$ & 0.15 & 0.15 & 0.18 & 0.20 & 0.24 & 0.28 & 0.30 & 0.28 & 0.28 & 0.20 & 0.20 \\ 
                $q_{0.84}$ & 0.34 & 0.30 & 0.32 & 0.35 & 0.38 & 0.38 & 0.39 & 0.38 & 0.38 & 0.38 & 0.28 \\ 
                $q_{0.975}$ & 0.40 & 0.40 & 0.40 & 0.40 & 0.40 & 0.40 & 0.40 & 0.40 & 0.40 & 0.40 & 0.36 \\ 
                \hline\hline
                \multicolumn{12}{c}{$M_2 > 1.5$ $M_{\sun}$}\\
                \hline
                \multicolumn{12}{c}{unknown mixing-length value}\\
                $q_{0.025}$ & 0.00 & 0.00 & 0.00 & 0.00 & 0.00 & 0.00 & 0.00 & 0.00 & 0.00 & 0.00 & 0.10 \\ 
                $q_{0.16}$ & 0.03 & 0.03 & 0.03 & 0.02 & 0.03 & 0.03 & 0.03 & 0.03 & 0.03 & 0.07 & 0.15 \\ 
                $q_{50}^{l}$ & 0.13 & 0.15 & 0.14 & 0.10 & 0.12 & 0.17 & 0.17 & 0.17 & 0.11 & 0.16 & 0.20 \\ 
                $q_{50}^{u}$ & 0.17 & 0.17 & 0.16 & 0.20 & 0.21 & 0.20 & 0.20 & 0.20 & 0.17 & 0.21 & 0.23 \\ 
                $q_{0.84}$ & 0.33 & 0.30 & 0.31 & 0.35 & 0.38 & 0.38 & 0.38 & 0.38 & 0.38 & 0.33 & 0.30 \\ 
                $q_{0.975}$ & 0.40 & 0.40 & 0.40 & 0.40 & 0.40 & 0.40 & 0.40 & 0.40 & 0.40 & 0.40 & 0.39 \\ 
                \hline
                \multicolumn{12}{c}{unknown element diffusion efficiency}\\
                $q_{0.025}$ & 0.00 & 0.00 & 0.00 & 0.00 & 0.00 & 0.00 & 0.00 & 0.00 & 0.00 & 0.00 & 0.10 \\ 
                $q_{0.16}$ & 0.04 & 0.05 & 0.03 & 0.03 & 0.03 & 0.03 & 0.03 & 0.03 & 0.03 & 0.05 & 0.15 \\ 
                $q_{50}^{l}$ & 0.15 & 0.15 & 0.14 & 0.20 & 0.21 & 0.19 & 0.17 & 0.20 & 0.11 & 0.17 & 0.18 \\ 
                $q_{50}^{u}$ & 0.17 & 0.16 & 0.20 & 0.24 & 0.27 & 0.28 & 0.33 & 0.30 & 0.30 & 0.17 & 0.20 \\ 
                $q_{0.84}$ & 0.33 & 0.30 & 0.35 & 0.38 & 0.38 & 0.38 & 0.39 & 0.38 & 0.38 & 0.35 & 0.25 \\ 
                $q_{0.975}$ & 0.40 & 0.40 & 0.40 & 0.40 & 0.40 & 0.40 & 0.40 & 0.40 & 0.40 & 0.40 & 0.35 \\ 
                \hline
        \end{tabular}
\end{table*}

To explore in greater detail the bias that is due to the lack of information about the correct mixing-length value to adopt, we present in Table~\ref{tab:bias-ml-M2} the cumulated envelope obtained by considering only systems with a secondary star more massive than 1.4 $M_{\sun}$ and than 1.5 $M_{\sun}$. These stars have a progressively thinner convective envelope, and therefore they are increasingly more independent of the efficiency of the superadiabatic convection.
As expected, in these subsets of masses the width of the bias interval in $\beta$, for $r = 1.0$ shrinks from 0.10 in the standard scenario to 0.05 and 0.03, respectively.

\subsection{Helium and heavy element diffusion}
\label{sec:diff}

Another tricky process to implement in stellar evolution codes is the diffusion of 
helium and heavy elements. Depending on the efficiency of microscopic diffusion, the 
surface abundances of the chemical elements change with time.  As an example, during the central hydrogen burning, 
the surface [Fe/H] drops from the ZAMS value and reaches a minimum at about 90\% of its evolution before central hydrogen
depletion \citep[see e.g. Fig.~12 in][]{scepter1}. After this point the convective envelope sinks inwards in more internal regions, 
where metals were previously accumulated by gravitational settling, leading to an increase of the surface metallicity. 
The size of the effect depends on both the mass of the star and its initial metallicity. The lower the initial metallicity and
the thinner the convective envelope, the higher the surface metallicity drop.

In the considered mass range of MS stars the observed [Fe/H] might therefore be quite different from the initial one -- if inhibiting processes are inefficient -- 
 depending on the stellar age and mass. Hence the microscopic diffusion effect should be taken into account whenever 
 the structural characteristics of low-mass MS stars have to be determined.
Neglecting microscopic diffusion in the stellar models used in the recovery procedure and assuming that the observed [Fe/H] is coincident with the initial one will introduce a systematic bias.

However, the efficiency of diffusion has been questioned by some authors  \citep[see e.g.][]{Korn2007, Gratton2011, Nordlander2012, Gruyters2014}.
Moreover, some widely used stellar model grids in the literature, namely BaSTI
\citep{teramo04,teramo06} and STEV \citep{padova08,padova09}, do not implement
diffusion. The same occurs in stellar models used in some grid-based
techniques, such as RADIUS \citep{Stello2009} and SEEK, which both adopt a grid of
models computed with the Aarhus STellar Evolution Code \citep{Dalsgaard2008}.

Thus, it is worthwhile to analyse the distortion in the inferred value of the convective-core overshooting efficiency
 that arises whenever the effects of the diffusion are neglected in stellar models.
The maximum bias caused by the element diffusion uncertainty on the estimated $\beta$ value was assessed by computing a grid of models with $\beta = 0.2$ and no element diffusion. As in the previous section, a synthetic dataset of $N = 50\,000$ binary systems was sampled from this grid and subjected to random perturbations. The $\beta$ values were then estimated by adopting the standard multi-overshooting grid. 

\begin{figure}
        \centering
        \includegraphics[height=8.5cm,angle=-90]{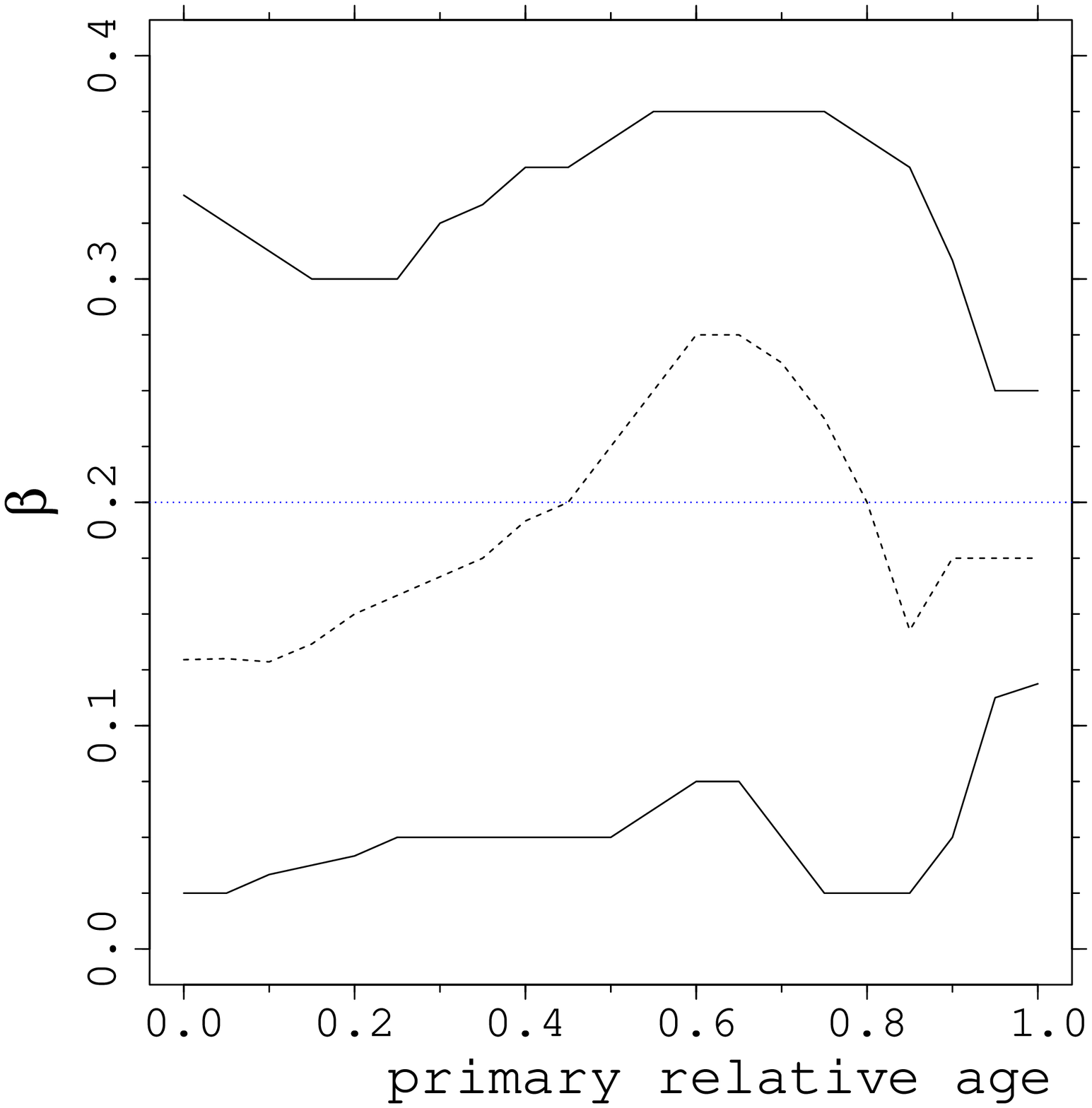}
        \caption{Same as the right panel of Fig.~\ref{fig:envelope-std}, but sampling from a grid without element diffusion.
        }
        \label{fig:envelope-nd}
\end{figure}

\begin{figure*}
        \centering
\includegraphics[height=17cm,angle=-90]{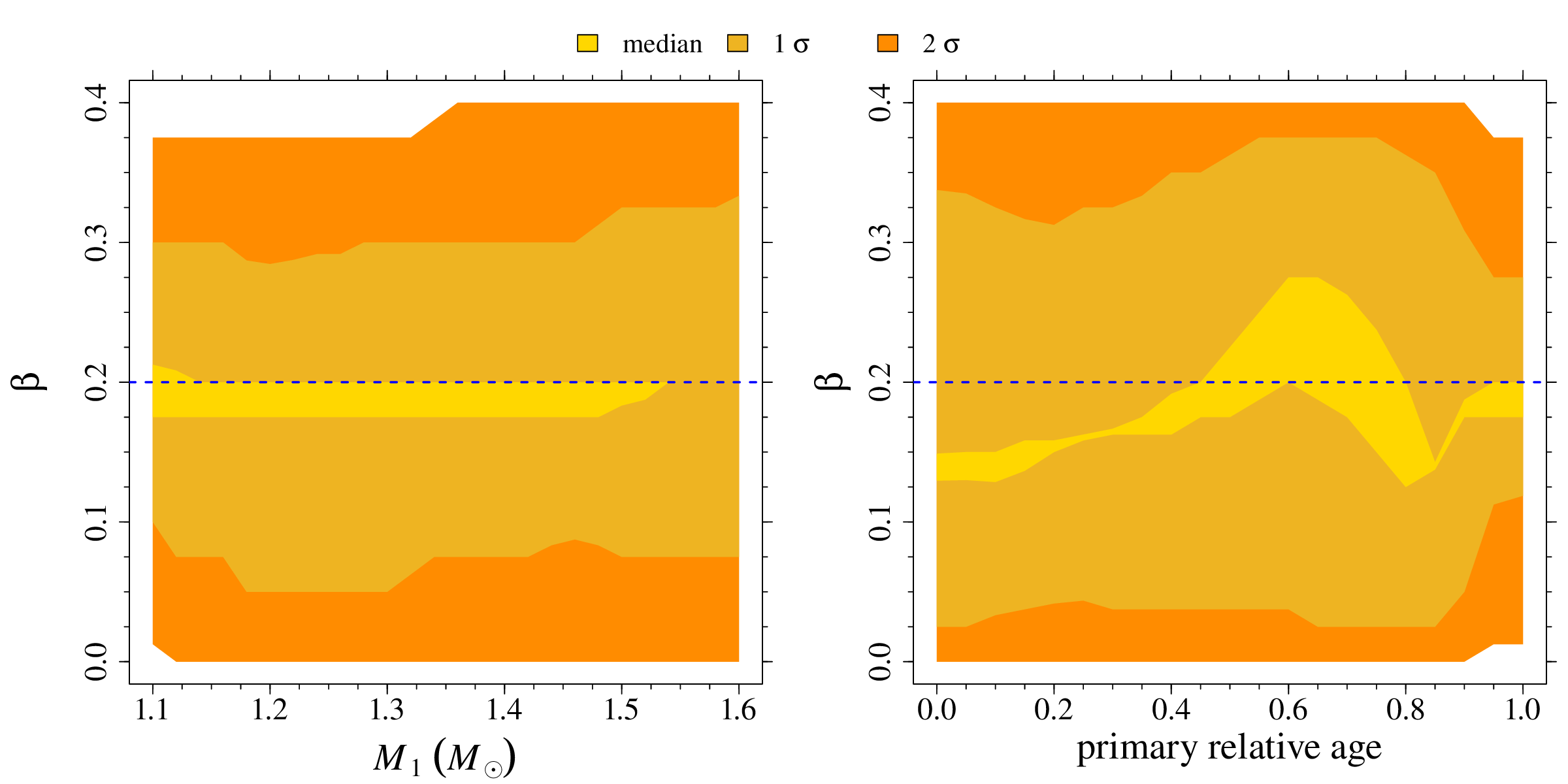}
\caption{Same as Fig.~\ref{fig:envelope-He-col}, but for the scenario without diffusion.}
\label{fig:envelope-nd-col}
\end{figure*}

The results are presented in Fig.~\ref{fig:envelope-nd}; the figure shows the $1 \sigma$ $\beta$ envelope as a function of the relative age of the primary star. Overall, the diffusion-induced bias is the lowest of those explored in the paper. The largest difference from the standard scenario occurs for 
systems whose primary star has a relative age in the range $0.5 < r < 0.8$. In the terminal phase of the evolution, a small underestimation of 0.025 occurs.   
The final uncertainty on $\beta$ that is due to the lack of knowledge of diffusion efficiency is presented in Fig.\ref{fig:envelope-nd-col} and Tables~\ref{tab:bias-he-ml-M} and \ref{tab:bias-he-ml-r}. 
As for the mixing-length variability, we present in
Table~\ref{tab:bias-ml-M2} the detail about the expected bias and random uncertainty in the subsets of masses obtained by restricting to systems with secondary star more massive than 1.4 $M_{\sun}$ and than 1.5 $M_{\sun}$. 
As a result of two competing effects (i.e. the thinning of the convective envelope and the faster evolutionary timescale), we obtained that the differences among all these scenarios are almost negligible.

\section{Conclusions}\label{sec:conclusions}

We theoretically investigated the statistical and systematic errors in the calibration of the convective-core overshooting efficiency on low-mass binary systems whose two components are in the MS phase. 
We used the grid-based pipeline SCEPtER \citep{scepter1,eta,binary}
and adopted
as observational constraints the stellar effective temperature, the metallicity [Fe/H], the mass, and the radius of the two stars.
The grid of stellar models was computed for the evolutionary phases
from the ZAMS to the central hydrogen depletion in the mass
range [1.1; 1.6] $M_{\sun}$, with 17 different values of the core overshooting parameter $\beta$ from 0.0 to 0.4.

The statistical uncertainty that affects the  $\beta$ value provided by the calibration procedure arising from the observational errors is very large. Moreover, the calibrated parameter was generally biased, the exact magnitude depending on the evolutionary phase of the stars and on the assumed true value of $\beta$.  As reference scenario we adopted a mild overshooting $\beta = 0.2$. In this case only for systems with primary star in the last 5\% of the MS evolution, the estimate is unbiased with a $1 \sigma$ random error ranging from  $+0.08$ to $-0.03$. 
Earlier evolutionary stages show a biased estimate of about $-0.04$ and a much higher variability; for binary systems whose primary star has a relative age $r \leq 0.8,$ the $1 \sigma$ envelope is practically unconstrained in the whole range of $\beta$ [0.0; 0.4].
The random  $1 \sigma$ uncertainty affecting the estimated $\beta
 $ is weekly dependent on 
the reference value used to build the synthetic dataset. 

Interestingly, in the scenario where the synthetic stars do not take  core overshooting into account (i.e. true value of $\beta = 0.0$), the calibration procedure is always biased because it always provides a very mild overshooting 
($+0.05$ at late evolutionary phases and $+0.12$ before).
This constitutes a serious problem for calibration studies, and fitted values of $\beta \leq 0.1$ should be considered with caution.  
         
In addition to the effect that is due only to the current uncertainties that affect the observables used in the estimate procedure,
we also studied the theoretical biases caused by still poorly contained parameters that affect stellar model behaviour. 
We focused on the initial helium abundance, the mixing-length value, and the efficiency of element diffusion. The first source of uncertainty is also relevant for more massive objects and  prevents a firm determination of $\beta$. By assuming an uncertainty of $\pm 1$ in the helium-to-metal enrichment ratio $\Delta Y/\Delta Z$ required to determine the initial helium content used to compute the evolution of the artificial stars, we found a large systematic uncertainty  on the $\beta$ value for the first 85\% of the evolution of the primary star; the maximum bias reaches 0.2 at $r = 0.6$. In the terminal phases of the evolution the effect of the initial helium uncertainty is of about 0.03. 
Taking  both systematic and $1 \sigma$ statistical uncertainty
into account, we found that  the error on the estimated $\beta$ values ranges from $-0.05$ to $+0.10$ in the last part of the evolution. At $2 \sigma$ level, the $\beta$ values is practically unconstrained in the whole explored range of convective-core overshooting efficiency.
The results are significantly poorer for lower $r$. For $r \leq 0.8$ the $1 \sigma$ statistical uncertainty is basically unconstrained throughout the explored range of $\beta$. 

The lack of constraints on the mixing-length value is an important bias source in the explored mass range. We quantified the effect of an uniform variation of $\pm 0.24$ in the value of $\alpha_{\rm ml}$ used to compute the evolution of the synthetic stars. The largest bias occurs for binary systems whose primary star is in the last part of its MS evolution ($r \geq 0.9$) with an error on the estimated median $\beta$ from $-0.03$ to $+0.07$. The $1 \sigma$ uncertainty that addresses statistical and systematic error sources ranges from $-0.09$ to $+0.15$. 

Another uncertainty source in the considered mass range is the efficiency of microscopic diffusion. We quantified the effect of completely neglecting diffusion on the stellar evolution. The bias is smaller than those due to the other considered sources.
In this case, the $1 \sigma$ uncertainty that addresses statistical and systematic error sources ranges from $-0.08$ to $+0.08$.
 
In summary, the results presented in this study suggest that the calibration of the convective-core overshooting efficiency based on the observed mass, effective temperature, radius, and metallicity of binary systems -- whose components are in the MS phase and in the explored mass range -- is not statistically grounded. The random uncertainty affecting the calibrated $\beta$ value is so large that it undermines the possibility of providing it in a reliable way from a single (or a few) binary systems. Moreover, the highlighted systematic biases also shed some doubts on the possibility of adopting a set of binary stars to reduce the uncertainty on the recovered $\beta$.  However, in the near future the scenario could change because high-quality asteroseismic observations of binary stars will become available. These data will shed light on the internal structure of the stars, allowing a better constraint of the main parameters that govern the stellar evolution. 

The effectiveness of core overshooting calibration on binary stars in more advanced evolutionary phases and/or different mass range from those studied here deserves further investigations.

\begin{acknowledgements}
We warmly thank Derek Homeier for the detailed comments and suggestions that helped us very much in clarifying and improving this paper.
This work has been supported by PRIN-MIUR 2010-2011 ({\em Chemical and dynamical evolution 
        of the Milky Way and Local Group galaxies}, PI F. Matteucci),  PRIN-INAF 2012 ({\em The M4 Core Project with Hubble Space Telescope}, PI
L. Bedin ), and PRIN-INAF 2014 (\emph{The kaleidoscope of stellar populations in globular clusters with Hubble Space Telescope}, PI S. Cassisi). 
\end{acknowledgements}

\bibliographystyle{aa}
\bibliography{biblio}

\end{document}